\documentclass[manuscript]{aastex}

\usepackage{epsfig}

\shortauthors{Zhang et al.} \shorttitle{Helicity of Self-Similar Force-free Fields}

\begin{document}

\title{Magnetic Helicity of Self-Similar Axisymmetric Force-free Fields}

\author{Mei Zhang\altaffilmark{1,2}, Natasha Flyer\altaffilmark{3} \& Boon Chye Low\altaffilmark{2}}

\altaffiltext{1}{Key Laboratory of Solar Activity, National Astronomical Observatory, Chinese Academy of Sciences, Datun Road A20,
Chaoyang District, Beijing 100012, China}
\altaffiltext{2}{High Altitude Observatory, National Center for Atmospheric Research, PO Box 3000, Boulder, CO 80307, USA}
\altaffiltext{3}{Institute for Mathematics Applied to Geosciences, National Center for Atmospheric Research, PO Box 3000, Boulder,
CO 80307, USA}

\begin{abstract}
In this paper we continue our theoretical studies on addressing what are the possible consequences of magnetic helicity accumulation in the solar corona. Our previous studies suggest that coronal mass ejections (CMEs) are natural products of coronal evolution as a consequence of magnetic helicity accumulation and the triggering of CMEs by surface processes such as flux emergence also have their origin in magnetic helicity accumulation. Here we use the same mathematical approach to study the magnetic helicity of axisymmetric power-law force-free fields, but focus on a family whose surface flux distributions are defined by self-similar force-free fields. The semi-analytical solutions of the axisymmetric self-similar force-free fields enable us to discuss the properties of force-free fields possessing a huge amount of accumulated magnetic helicity. Our study suggests that there may be an absolute upper bound on the total magnetic helicity of all bipolar axisymmetric force-free fields. And with the increase of accumulated magnetic helicity, the force-free field approaches being fully opened up, with Parker-spiral-like structures present around a current-sheet layer as evidence of magnetic helicity in the interplanetary space. It is also found that among the axisymmetric force-free fields having the same boundary flux distribution, the one that is self-similar is the one possessing the maximum amount of total magnetic helicity. This possibly gives a physical reason why self-similar fields are often found in astrophysical bodies, where magnetic helicity accumulation is presumably also taking place.
\end{abstract}



\section{Introduction}

Magnetic helicity is a physical quantity that describes field topology, quantifying the twist (that is, self-helicity) and linkage (that is, mutual-helicity) of magnetic flux lines. One important property of magnetic helicity is that the total magnetic helicity is still conserved in the corona even when there is a fast magnetic reconnection (Berger 1984). This indicates that once the magnetic helicity is transported into the corona, it cannot be annihilated by solar flares.

Over the past two decades, observations have shown that magnetic fields, created by dynamo processes in the solar interior, are emerging at the solar photosphere and then into the corona, with a preferred helicity sign in each hemisphere, namely, positive helicity sign in the southern hemisphere and negative helicity sign in the northern hemisphere (Pevtsov et al. 1995, Bao \& Zhang 1998, Hagino \& Sakurai 2004, Zhang 2006, Wang \& Zhang 2010, Hao \& Zhang 2011). Taking this observed hemispheric rule of helicity sign, together with the approximate law of helicity conservation in turbulent magnetic reconnection (Berger 1984), we can then infer that the total magnetic helicity is accumulating in the corona in each hemisphere (Zhang \& Low 2005).

An interesting question then to address is, what are the consequences of this magnetic helicity accumulation in the corona.

In a previous paper (Zhang et al. 2006), we showed that in an open atmosphere such as the solar corona, for a given boundary flux distribution on the solar surface, there is an upper bound on the magnitude of the total magnetic helicity of all axisymmetric power-law force-free fields. The accumulation of magnetic helicity in excess of this upper bound would initiate a non-equilibrium situation, resulting in solar eruptions such as coronal mass ejections (CMEs), as natural products of coronal evolution due to the magnetic helicity accumulation in the corona.

In a follow-up paper (Zhang \& Flyer 2008), we studied the dependence of the helicity bound on the boundary condition. We found that the magnitude of the helicity upper bound of force-free fields is non-trivially dependent on the boundary flux distribution. Fields with a multipolar boundary condition can have a helicity upper bound ten times smaller than those with a dipolar boundary condition. This suggests that a coronal magnetic field may erupt into a CME when the applicable helicity bound falls below the already accumulated helicity as the result of a slowly changing boundary condition. This gives insights into the observed associations of CMEs with the magnetic variations on the surface. For example, it can explain why CME occurrences are often observed to be associated with flux emergences whereas not every flux emergence will trigger a CME (e.g. observations in Zhang et al. 2008). Here the role of flux emergence is modeled as a change in boundary condition. This change is not always a trigger of a CME, a role it can play only if the coronal magnetic field has accumulated a critical amount of magnetic helicity.

In this paper, we continue our study in this direction by investigating magnetic helicity that relates to a family of axisymmetric self-similar force-free fields. The semi-analytical solutions of this family enable us to discuss the properties of force-free fields possessing a huge amount of accumulated magnetic helicity; impossible otherwise using conventional numerical methods as in our previous studies. We are interested in addressing the following two questions: 1) is there an absolute helicity upper bound for all bipolar axisymmetric force-free fields? 2) when a huge amount of magnetic helicity is finally released into the interplanetary space, what would the magnetic field look like? We organize our paper as follows. We describe our theoretical model in Section 2 and give our results and analysis in Section 3. Conclusion and discussion are given in Section 4.


\section{The Model}

\subsection{Axisymmetric power-law force-free fields}

As in our previous studies (Zhang et al. 2006, Zhang \& Flyer 2008), we use force-free fields to represent coronal magnetic fields.

Force-free field describes an equilibrium coronal magnetic field by setting the Lorentz force to be zero:
\begin{equation}
\label{fff1}
( \nabla \times {\bf B} ) \times {\bf B} = 0 ~~,
\end{equation}
\noindent where ${\bf B}$ is the vector magnetic field, subject to the usual solenoidal condition:
\begin{equation}
\label{solenoid}
\nabla \cdot {\bf B} = 0 ~~.
\end{equation}

Being force-free is usually regarded as a good approximation for large-scale coronal magnetic fields. In the corona, the plasma is very tenuous. An equilibrium field of characteristic intensity 10G or larger, in the presence of current, is capable of exerting a Lorentz force much larger than can be balanced by typical coronal fluid forces.

Among the force-free fields, we further focus on a special family of axisymmetric power-law force-free fields to understand the basic physical properties of interest.

By assuming axisymmetry, the magnetic field {\bf B} in the corona ($r > 1$) is written as
\begin{equation}
{\bf B} = {1 \over r \sin \theta} \left[
{1 \over r}{\partial A \over \partial \theta} ~,
~ - {\partial A \over \partial r}~, ~Q (r, \theta) \right] ~,
\end{equation}
\noindent where the flux function $A$ defines the poloidal magnetic field and the function $Q(r, \theta)$ defines the azimuthal field component. The base of the corona has been taken as $r=1$ in spherical coordinates.

The total (relative) magnetic helicity of this field, derived in Zhang et al. (2006), is
\begin{equation}
H = 4 \pi \int_{r>1} A Q {d\theta \over \sin \theta} dr ~.
\end{equation}
\noindent Note that this formula applies to all axisymmetric fields, regardless of the form of $Q(r, \theta)$ or whether the field is force-free or not.

With magnetic field {\bf B} in the form of Eq. (3), Eq. (1) requires $Q(r, \theta)$ take the form of $Q(A)$. For mathematical tractability, we further assume $Q(A)$ takes a form of power law given by
\begin{eqnarray}
Q^2 (A) = \frac{2 \gamma}{n+1} A^{n+1} ~,
\end{eqnarray}
\noindent as in Flyer et al. (2004, 2005), Zhang et al. (2006) and Zhang \& Flyer (2008). Here $n$ is an odd constant index required to be no less than 5 in order for the field to possess finite magnetic energy in $r > 1$. $\gamma$ is a free parameter that we choose to be positive without loss of generality.

These power-law force-free fields can be taken as representatives of the much bigger set of complicated nonlinear force-free fields with arbitrary prescriptions of $Q(A)$. Note that we cannot force the real corona to follow such a restrictive form of $Q (A)$, particularly not that the corona must evolve along the fields described by these specialized fields. These fields are analyzed here, first because their solutions are mathematically tractable; secondly because their richness (solutions with different $n$ and $\gamma$ values) makes these solutions spread into a large, if not a full, space of physical quantities. Thus, they can be taken as good representatives of a full set of force-free fields to address a few physical questions. In particular, they allow us to explore the physics of the total magnetic helicity of force-free fields which is our main interest in this series of papers.

With this form of $Q(A)$, the force-free condition Eq. (1) is reduced to following single governing equation:
\begin{eqnarray}
{\partial^2 A \over \partial r^2} + {1 - \mu^2 \over r^2}
{\partial^2 A \over \partial \mu^2} + \gamma A^n = 0 ~~,\label{eqnA}
\end{eqnarray}
\noindent where $\mu=\cos\theta$.

\subsection{Axisymmetric self-similar force-free fields}

Among all the solutions to Eq.(6), there is a sub-family, first constructed in Low \& Lou (1990), and we will refer to them as axisymmetric self-similar force-free fields hereafter.

Self-similar solutions have been studied extensively in the past, either for addressing solar problems (e.g. Low 1982a, 1982b, 1984a, 1984b, 1992) or for attacking astrophysical questions (e.g. Shu 1977, Shu et al. 2002). A time-dependent three-dimensional self-similar MHD model of CMEs has also been built (Gibson \& Low 1998) and it is found that its solutions can address a few observational phenomena rather well (e.g. Gibson \& Low 2000, Dove et al. 2011).

Axisymmetric self-similar force-free fields in Low \& Lou (1990) were first constructed to study the three-dimensional nonlinear force-free fields in Cartesian geometry. With a few chosen parameters the constructed fields give a geometry looking strikingly like active regions on the Sun, and yet all solutions are semi-analytical and so can be used to test various numerical methods for extrapolating coronal magnetic fields from the vector magnetic fields measured on the photosphere (e.g. Schrijver et al. 2006, Liu et al. 2011).

In order to produce axisymmetric self-similar force-free fields, we need to further assume that the flux function $A$ takes a form of
\begin{equation}
A (r, \mu) = \frac{P(\mu)}{r^{\frac{2}{n-1}}} ~~~,
\end{equation}
\noindent in addition to those assumptions made in the last subsection. Here $n$ takes the same notation as in Eq.(5). This approach separates the variation in $r$-direction from that in $\mu$-direction and makes the fields self-similar. By self-similar, we mean that the latitudinal variation of the flux function $A$, defined by $P(\mu)$, is identical in form at any $r$, except for its magnitude, given by $r^{\frac{2}{n-1}}$.

Now with Eq.(7), Eq.(6) further reduces to
\begin{eqnarray}
(1 - \mu^2){d^2 P \over d \mu^2} + \frac{2(n+1)}{(n-1)^2} P + \gamma P^{\frac{n+1}{2}} = 0 ~~,
\end{eqnarray}
\noindent with boundary conditions:
\begin{equation}
P = 0 ~~~~ at ~~~~ \mu = -1, 1 ~~~~.
\end{equation}

Solving Eq.(8) admits eigenfunction solutions $P(\mu)$ with associated eigenvalues of $\gamma$ and hence the axisymmtric self-similar force-free fields by Eq.(7). Because Eq.(8) is 1D, we can solve it with accuracy to very high $n$. In a sense, these solutions are semi-analytical. We numerically solve Eq.(8) for $n=5, 7, 9, 11, 13, 15, 17$ and so on, all the way to $n=201$. For each $n$, the field has been normalized to make the surface flux function at the equator, $A(r=1, \theta=90^\circ)$, equal to 1.

Note that hereafter, we denote the eigenvalues of $\gamma$ to Eq.(8) as $\gamma_{ss}$, to distinguish them from other $\gamma$ values of general solutions to Eq.(6). For each $n \ge 5$, there are more than one eigenvalue of $\gamma_{ss}$. We always choose the lowest eigenvalue to assure that our axisymmtric self-similar force-free fields are bipolar. Thus, for $n=5, 7, 9, 11, 13, 15, 17 ... 201$, we obtain 99 axisymmtric self-similar force-free fields and they are all bipolar.

\subsection{The solutions of flux function}

Solving the nonlinear equation Eq.(6) in a general form, even in 2D, is not a trivial undertaking. As before, we use the mathematical and numerical methods presented in Flyer et al. (2004) to solve Eq.(6). These methods include the Newton's iteration combined with a pseudo-arc-length continuation scheme to guarantee the completeness of each solution branch generated by the $\gamma$ values.

By solution branch, we mean that for any given boundary flux distribution and each $n$, we can obtain a complete set of different $\gamma$ values and hence different force-free fields and they form one solution branch. Note that for this nonlinear boundary value problem in 2D, solutions exist, with their $\gamma$ values bounded by a maximum value $\gamma_{max}$. This is not always the case when dealing with truly 3D magnetic fields. In 3D, continuous force-free fields are not guaranteed to exist if the boundary conditions are arbitrarily prescribed, as discussed in Low \& Flyer (2007).

In this study, we solve Eq.(6), by taking the boundary flux distributions as those of the axisymmtric self-similar solutions, that is, the eigenfunctions $P(\mu)$ defined by Eq.(8). Note that for each $n$, we have one particular $\gamma_{ss}$, the eigenvalue defined by Eq.(8), but we have a series of $\gamma$ values, defined by Eq.(6). In other words, for each $n$, the eigenvalue $\gamma_{ss}$ picks up the unique self-similar solution for that $n$ to fix the boundary flux distribution at $r=1$, and Eq.(6) gives other solutions of force-free fields that are not self-similar.

We obtain the axisymmetric power-law force-free fields, solutions to Eq.(6), for $n=5, 7, 9, 11, 13, 15, 17$ cases. Note that previously we can only solve Eq.(6) with $n=5, 7, 9$. Now with a development in our numerical method (Fornberg et al. 2012) we can solve Eq.(6) to $n=17$ in this study.

Figure 1 displays plots of the physical quantities, such as magnetic energy $E=\int_{r>1}\frac{B^2}{8\pi}dV$, total azimuthial flux $F_\varphi=\int_{r>1}|B_\varphi|rdrd\theta$ and total magnetic helicity $H$ as in Eq.(4), of all possible solutions with different $\gamma$ values for $n=5$ (top panels), $n=7$ (middle panels) and $n=9$ (bottom panels) fields. We call these curves solution curves. They are similar to those in Figure 2 of Zhang et al. (2006) and in Figure 2 of Zhang \& Flyer (2008). As we can see from the development of these curves, there seems to be an upper bound on the total magnetic helicity, as well as upper bounds on the total azimuthal flux and on the magnetic energy, for force-free fields with the same boundary flux distribution, similar to what we have found in previous studies.


\section{Results and Analysis}

\subsection{Self-similar force-free fields as maximum-helicity end-states}

In addition to confirming the existence of a helicity upper bound, another interesting phenomena observed from these numerical calculations is that the self-similar force-free field is the end-state on each solution curve with the same boundary flux distribution and same $n$.

In Figure 1, the dashed line in each panel shows the value of either the azimuthal flux (middle columns) or the total magnetic helicity (right columns) of each self-similar force-free field with its corresponding $n$, calculated separately using the semi-analytical solution described in Section 2.2 as well as in Low and Lou (1990). We see here that the evolutions of the azimuthal flux and the total magnetic helicity indeed end with the exact numbers of corresponding self-similar force-free fields.

Note that not only the azimuthal flux and the total magnetic helicity of the end-states approach to those of the self-similar force-free fields, the whole end-state field actually becomes identical to the self-similar force-free field in each solution curve, as shown in Figures 2, 3 and 4. In Figure 2, we show the results from the $n=5$ case calculations. The top panels show the field configuration and the flux function at the equator, i.e. $A(r, \theta=90^\circ)$, for the potential field with $H=0$, i.e. the starting point of the solution curve in Figure 1. In the middle panels, we show another field, which is located further along the solution curve in Figure 1 with more total magnetic helicity ($H=6.03$). In the bottom panels, we show the end-state in the solution curve, which has $H=8.62$. As a comparison, the semi-analytical $n=5$ self-similar force-free field is also plotted in dashed lines. We see that in the beginning (that is, the top panels) the field deviates from the self-similar force-free field significantly. With the development along the solution curve, the difference becomes smaller. When the field reaches the end-state along the solution curve, the difference between the numerical solution to Eq.(6) and the semi-analytical solution to Eq.(8) has disappeared. This means that the end-state becomes identical to the self-similar force-free field.

Similar observations can be found in Figure 3, where the $n=9$ case is shown, and in Figure 4, where the $n=13$ case is shown. This interesting observation tells us that the self-similar force-free field is the end-state that possesses the maximum total magnetic helicity among those force-free fields that have the same boundary flux distribution and the same $n$. We know that many astrophysical phenomena can be described by self-similar processes. We also know that magnetic helicity is accumulating in the solar corona with a possibility that similar helicity accumulation process also takes place in the corona of other astrophysical bodies. Our observations from the simple numerical model here give an indication to why self-similar processes are more common in nature than would have been suggested by their special mathematical solutions (Barrenblatt \& Zel'dovich 1972). It is their physical property that self-similar fields are equilibrium states with maximum total magnetic helicity that makes these fields to be found in nature, in spite of their specialized forms of solution.

\subsection{An upper bound on the total magnetic helicity for all bipolar axisymmetric force-free fields}

With an improvement in the continuation method in our numerical scheme (Fornberg et al. 2012) we can obtain numerical solutions to Eq.(6) for $n=5, 7, 9$, up to $n=17$. Beyond that, it becomes numerically too expensive to resolve the tiny bubble (flux ropes) in the process to reach the end-state along the solution curve, using the Chebyshev-Fourier discretization of Flyer et al. (2004). However, once we have a basis to that the self-similar force-free field is the end-state along each solution curve, we can then use the semi-analytical self-similar force-free fields to study how the magnitude of the upper bound on the total magnetic helicity evolves with the increase of $n$.

Figure 5 shows how the total magnetic helicity of the end-state in each solution curve, normalized by the square of corresponding surface poloidal flux ($F_p^2$) as in Zhang \& Flyer (2008), evolves with the increase of $n$. Each plus symbol gives the magnitude of the total magnetic helicity of the self-similar force-free field and each diamond symbol shows the magnitude of the total magnetic helicity of the end-state field obtained numerically along each solution curve for $n=5, 7, 9$ up to $n=17$ case. We see here that as $n$ increases, the total magnetic helicity ($H/F_p^2$) in these force-free fields levels off, to an asymptotic value close to 0.5. This indicates that there may be an absolute helicity upper bound for all bipolar axisymmetric force-free fields, independent of the flux distribution, and that bound on the total magnetic helicity ($H$) is close to $0.5F_p^2$.

\subsection{Parker-spiral structures in the fully-open field as evidence of magnetic helicity in the interplanetary space}

As $n$ increases, not only does the total magnetic helicity in the self-similar force-free fields levels off, but also the magnetic field becomes more and more open. Figure 6 shows the evolution of the field configuration in a few representative self-similar force-free fields. We see that as $n$ increases, more and more field lines become open. When $n$ reaches 201, the field lines, projected on the $r-\theta$ plane, become almost everywhere radial, approaching to those in a fully opened-up field.

Figure 7 gives a comparison of the $n=201$ self-similar force-free field with the fully-open Aly field (Aly 1991). The Aly field defines a magnetic field that is fully opened up, with a current sheet at the equator if the field is axisymmetric as in our cases. The mathematical construction of the fully-open Aly fields can be found in Zhang \& Low (2001). The key point to mention here is that these fully-open Aly fields are actually potential fields, even though they possess significant amounts of magnetic energy because of the existence of the current sheet. Here we see that the field configurations in the $r-\theta$ plane of the $n=201$ self-similar force-free field and the fully-open Aly field are almost identical. From the bottom panel of Figure 7, we see also that with the increase of $n$, the magnetic energies of these two types of fields, the self-similar force-free fields and the fully-open Aly fields, also approach one another.

This is consistent with what Wolfson (1995) found in his numerical calculations. He used the same series of self-similar force-free fields constructed in Low and Lou (1990). The difference is that he interpreted the opening up of the field lines (such as what we show in Figure 6) as a result of the increase of shearing at the base of the corona. He also found that with the increase of the shearing (equivalent to increasing $n$ in our discussions) the field becomes fully opened up and the magnetic energy gets close to the Aly-limit (that is, the magnetic energy of the fully-open Aly field) as we have shown in Figure 7. So he concluded that, as the limiting case is approached, the azimuthal component vanishes, the field lines are truly radial and the field becomes indistinguishable from the case of the field having been opened without any shearing. In summary, he concluded that the limiting case (he used the $n=201$ self-similar case too) is indistinguishable from the fully-open Aly field.

Here we show that the limiting case of self-similar force-free fields, using $n=201$ case as an example, and the fully-open Aly field are actually distinguishable. First of all, the limiting case of self-similar force-free fields possesses a huge amount of magnetic helicity, with a number close to $0.5F_p^2$ that possibly sets the absolute upper bound on total magnetic helicity for all bipolar axisymmetric fields, whereas the fully-open Aly field contains zero magnetic helicity.

Secondly, even though the limiting case of self-similar force-free fields and the fully-open Aly field have a similar magnitude of magnetic energy, their vector magnetic fields at the base of the corona are evidently different. Figure 8 gives a comparison between the three components of the vector magnetic field at $r=1$, of the $n=201$ self-similar force-free field and the fully-open Aly field. We see that, they do have similar normal magnetic field distributions ($B_r$) and the similar surface integrands (that is, the function of $B_r^2-B_t^2-B_v^2$ at $r=1$), where the latter results in a close number of magnetic energy of the two fields. However, the distributions of their transverse magnetic field ($B_t$) and of the azimuthal magnetic field ($B_v$) are significantly different.

Further differences of the three components between the two magnetic fields can be seen in Figure 9. Here the variations of flux function, normal magnetic field $B_r$, transverse magnetic field $B_t$ and azimuthal magnetic field $B_v$ at the equator, are plotted against the radial distance $r$. We see that for the Aly field, the flux function at the equator does not drop with the increase of $r$, whereas the flux function of the $n=201$ self-similar force-free field decreases with $r$. We see also that there is an azimuthal magnetic field $B_v$ at the equator for the $n=201$ self-similar force-free field, but for the Aly field, $B_v=0$ by definition. The variations with $r$ of $B_r$ and $B_t$ at the equator are also different for the $n=201$ self-similar force-free field and the Aly field. So, these two fields are actually quite different.

Finally, we show that even though their field geometries in the $r-\theta$ plane look almost identical (that is, the top panels in Figure 7), their 3D field line structures are actually quite different. Figure 10 shows a comparison of the 3D structures of field lines between the $n=201$ self-similar force-free field (purple lines) and the potential fully-open Aly field (blue lines). From top to bottom panels, plotted respectively are the field lines $0.5^\circ$, $1^\circ$, $2^\circ$ and $20^\circ$ away from the equator. The equator is located in the X-Y plane and the poles in the Z direction. The length of each axis in the left panels is 100 solar radius and the length of each axis in the right panels is 5 solar radius. The central red sphere shows the size of the Sun. We see that, whereas the field lines in the Aly field are all radial, the field lines in the $n=201$ self-similar force-free field show impressive spiral structures, a version of the Parker spirals. This is quite understandable because in one case the field is potential and has no magnetic helicity, whereas in the other case there is a huge amount of total magnetic helicity stored in the field. The field lines between these two fields become undistinguishable only if they originate from latitudes larger than $20^\circ$ from the equator (that is, where the current sheet lies in). Due to these results, we speculate that the existence of Parker-spiral-like structures in the field is the way how the magnetic field in the interplanetary space possesses a huge amount of magnetic helicity and yet, at the same time, keeps all its field lines fully open to infinity.


\section{Conclusion and discussion}

In our previous papers, we studied series of axisymmetric power-law force-free fields and found that there may be an upper bound on the total magnetic helicity that force-free fields with the same boundary flux distribution can contain (Zhang et al. 2006) and this upper bound non-trivially depends on the boundary flux distribution (Zhang \& Flyer 2008). These studies put solar eruptions such as CMEs as natural products of coronal evolution and give physical understandings on why surface variations such as flux emergences can trigger CMEs.

In this paper, we continue our studies in this direction with further analysis on the series of solutions to Eq.(6). We make use of a family of semi-analytical self-similar force-free fields constructed in Low and Lou (1990) to extend the force-free fields we studied beyond what our current numerical methods can obtain. The results we find are summarized below.

1. Within the same family of axisymmetric power-law force-free fields that have the same boundary flux distribution and the same power-law index $n$, the self-similar force-free field is always the end-state that possesses the maximum total magnetic helicity. This means that self-similar force-free fields are good magnetic helicity containers. It also gives an indication to why self-similar fields are often found in the astrophysical systems, where magnetic helicity accumulation is presumably also taking place.

2. With the increase of the index n, the total magnetic helicity of self-similar force-free fields levels off to an asymptotic value. This number, $0.5F_p^2$, possibly defines the absolute upper bound on the total magnetic helicity that all bipolar axisymmetric force-free fields can contain, independent of the boundary flux distribution.

3. With the increase of the index n, the magnetic field fully opens up, forming a current sheet at the equator and reaches the Aly-limit energy as Wolfson (1995) has described. However, different from his conclusions, our study shows that even though the field geometry in the $r-\theta$ plane of this limiting case of self-similar force-free fields looks almost identical to the potential fully-open Aly field, the two fields are not topologically the same in 3D space. The limiting case of self-similar force-free field possesses structures with Parker spirals and contains a huge amount of magnetic helicity. This seems suggesting that Parker-spiral-like structures are where the significant amount of magnetic helicity, released from the low solar corona, resides in the interplanetary space.

Our finding that the limiting case of self-similar force-free fields is not the fully-open Aly field removes the concerns in Wolfson (1995). In fact, it actually strengthens the main result in his paper, i.e. the corona can become fully opened up by shearing the footpoints of its magnetic field, rather than weakening it. It is also interesting to mention that this process of opening up by shearing the footpoints in a particular manner, is another possible way to open up the magnetic field without magnetic reconnection, a theoretical phenomena described in Rachmeler et al. (2009).


\acknowledgements

We thank Piotr Smolarkiewicz for helpful comments. Mei Zhang acknowledges supports of the National Natural Science Foundation of China (Grants No. 10921303 and No. 11125314), the Knowledge Innovation Program of the Chinese Academy of Sciences (Grant No. KJCX2-EW-T07) and the National Basic Research Program of MOST (Grant No. 2011CB811401). Natasha Flyer would like to acknowledge the support of NSF grants ATM-0620100 and DMS-0934317. The National Center for Atmospheric Research (NCAR) is sponsored by the National Science Foundation.



\begin{figure}
\centerline{\includegraphics[width=120mm]{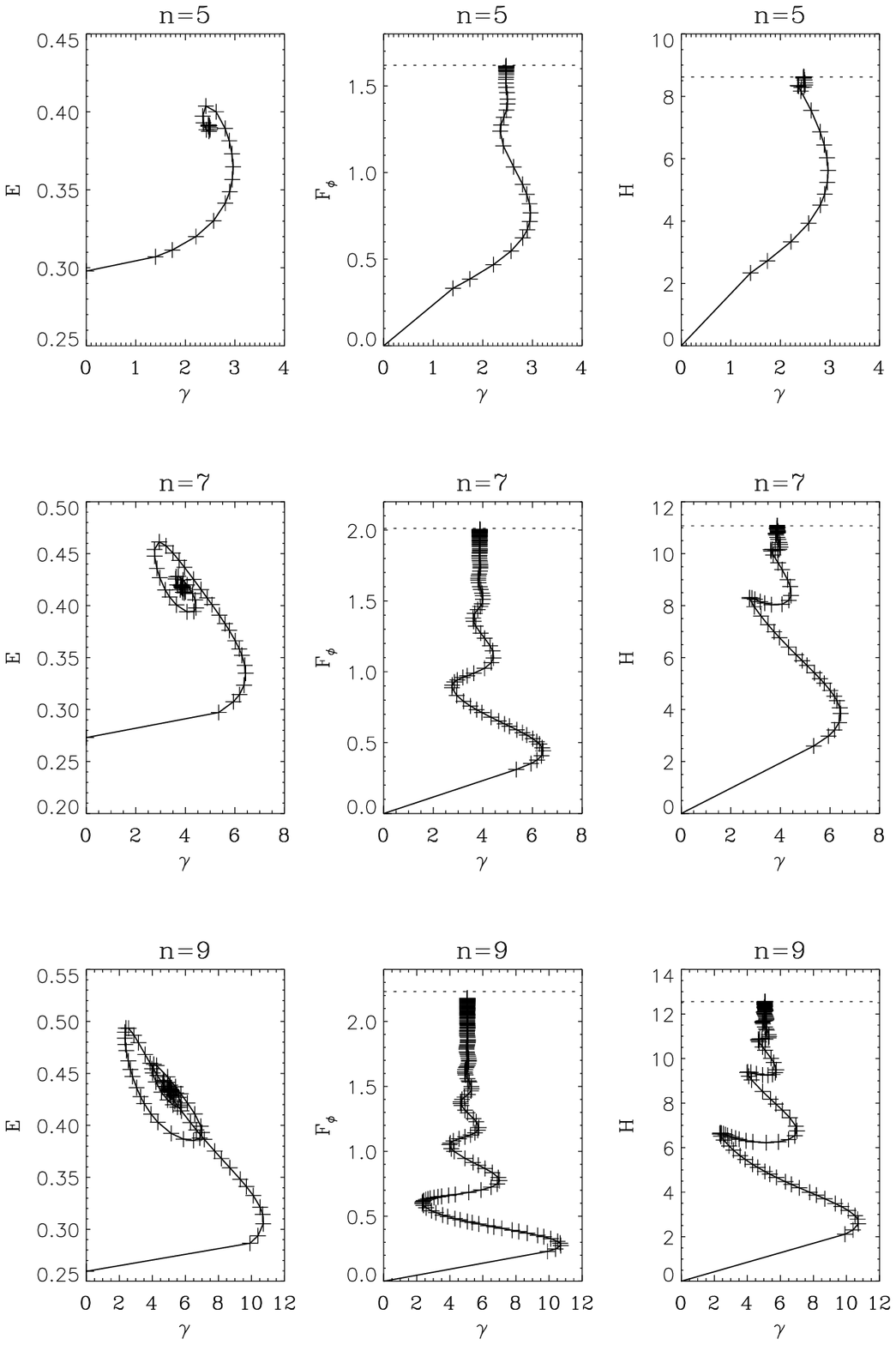}}
\caption{\small{Variation of total magnetic energy ($E$), azimuthal flux ($F_{\varphi}$) and total magnetic helicity ($H$) vs $\gamma$ along the solution curve for $n=5$ (top panels), $n=7$ (middle panels) and $n=9$ (bottom panels) fields. Each point, denoted by a plus symbol in the plot, represents a solution to Eq.(6). The dashed line in each panel shows the value of corresponding self-similar force-free field, calculated separately using the semi-analytical solution in Low and Lou (1990).}}
\end{figure}

\begin{figure}
\centerline{\includegraphics[width=120mm]{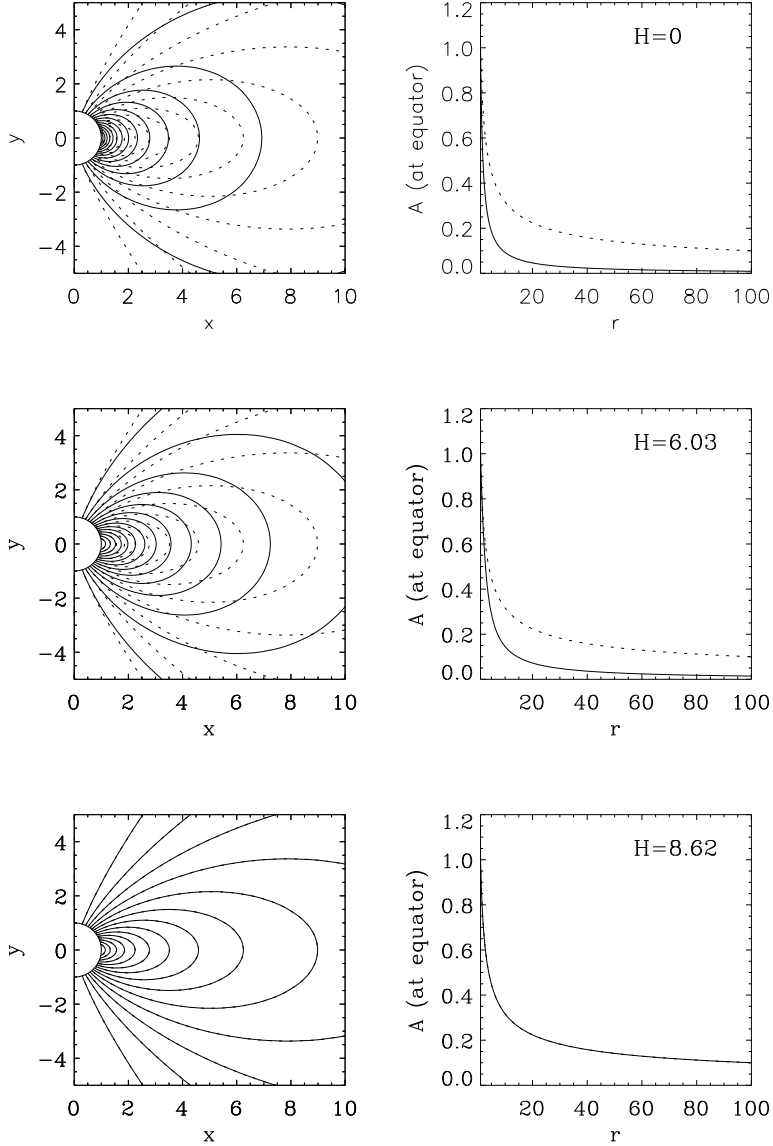}}
\caption{\footnotesize{Left panels: Field configurations of three representative fields selected from the $n=5$ solution curve. These contours of flux function represent the lines of force of the axisymmetric field projected on the $r-\theta$ plane, using Cartesian coordinates $(x,y)$ defined with on that plane. From top to bottom panels, each representative field possesses different amount of total magnetic helicity. Right panels: Variations of flux function at the equator, of each representative field respectively, vs the radial distances in the unit of solar radius. The solid lines in these panels are for the axisymmetric power-law force-free fields, and the dashed lines are for the self-similar force-free field with the same $n$.}}
\end{figure}

\begin{figure}
\centerline{\includegraphics[width=120mm]{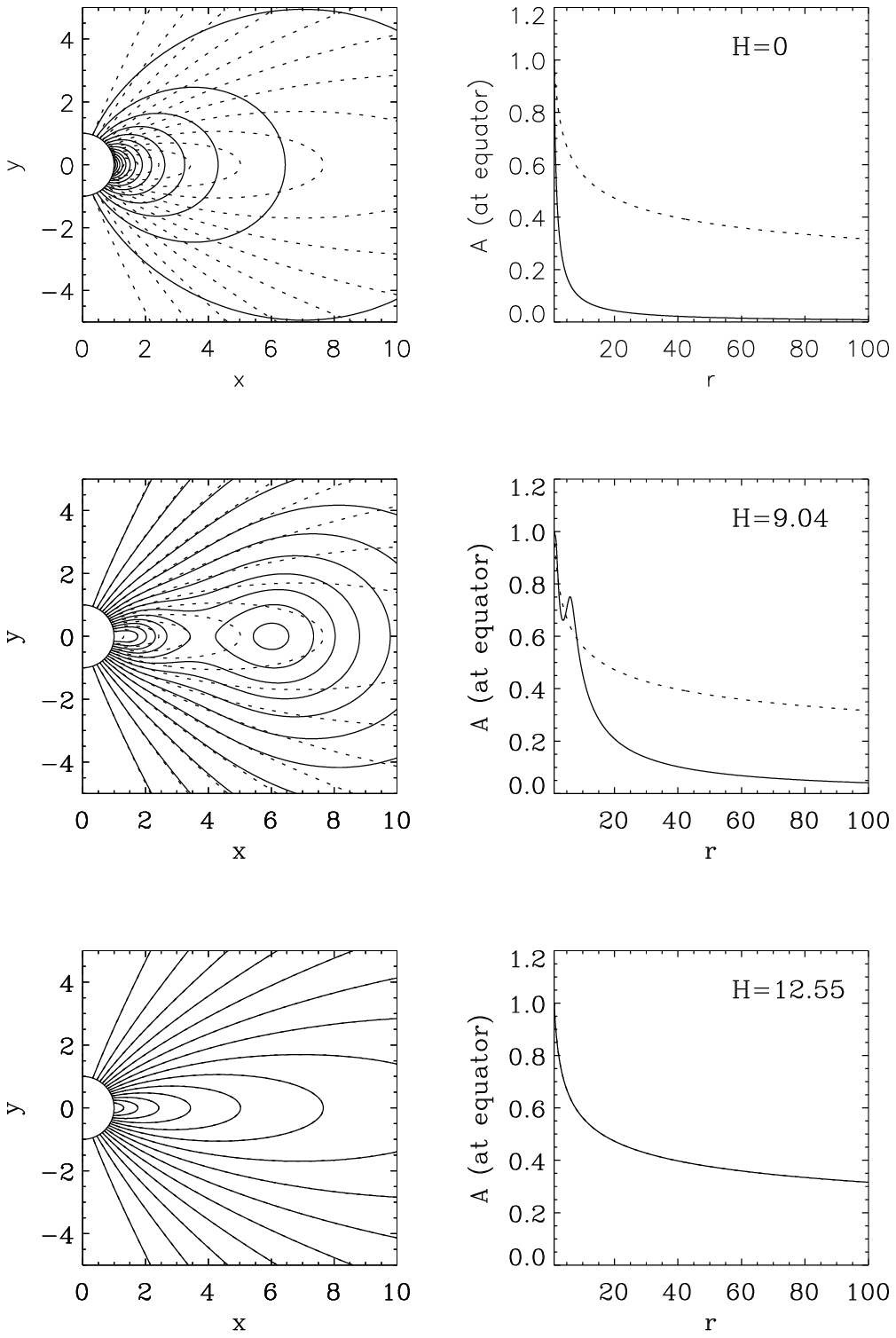}}
\caption{\small{Same as Figure 2, but for the $n=9$ case.}}
\end{figure}

\begin{figure}
\centerline{\includegraphics[width=120mm]{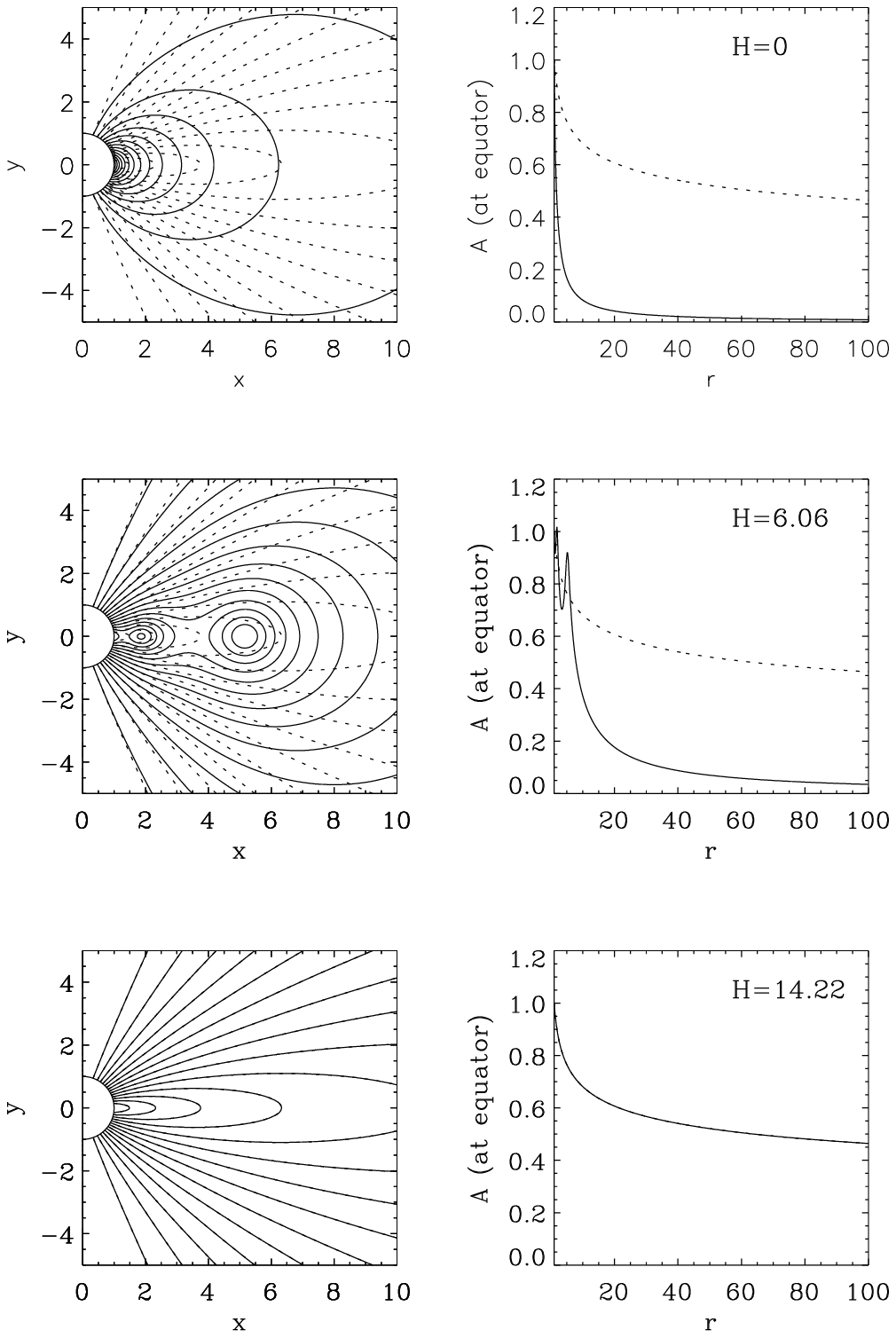}}
\caption{\small{Same as Figure 2, but for the $n=13$ case.}}
\end{figure}

\begin{figure}
\centerline{\includegraphics[width=120mm]{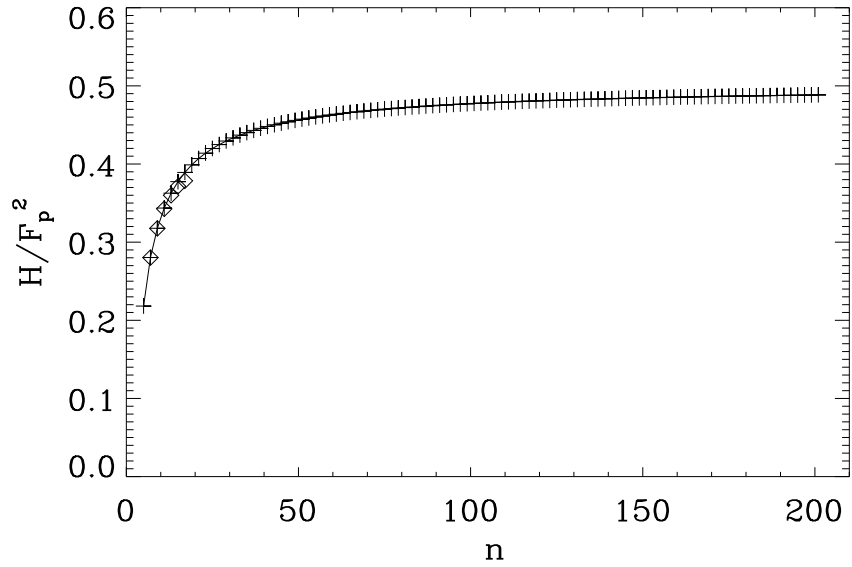}}
\caption{\small{Variation of the total magnetic helicity, normalized by the square of corresponding surface poloidal flux, with $n$. Plus symbols give the total magnetic helicity in self-similar force-free fields and the diamond symbols in end-state fields along solution curves with $n$ from 5 to 17. We see here that as $n$ increases, the total magnetic helicity ($H/F_p^2$) in these force-free fields levels off to an asymptotic value close to 0.5.}}
\end{figure}

\begin{figure}
\centerline{\includegraphics[width=130mm]{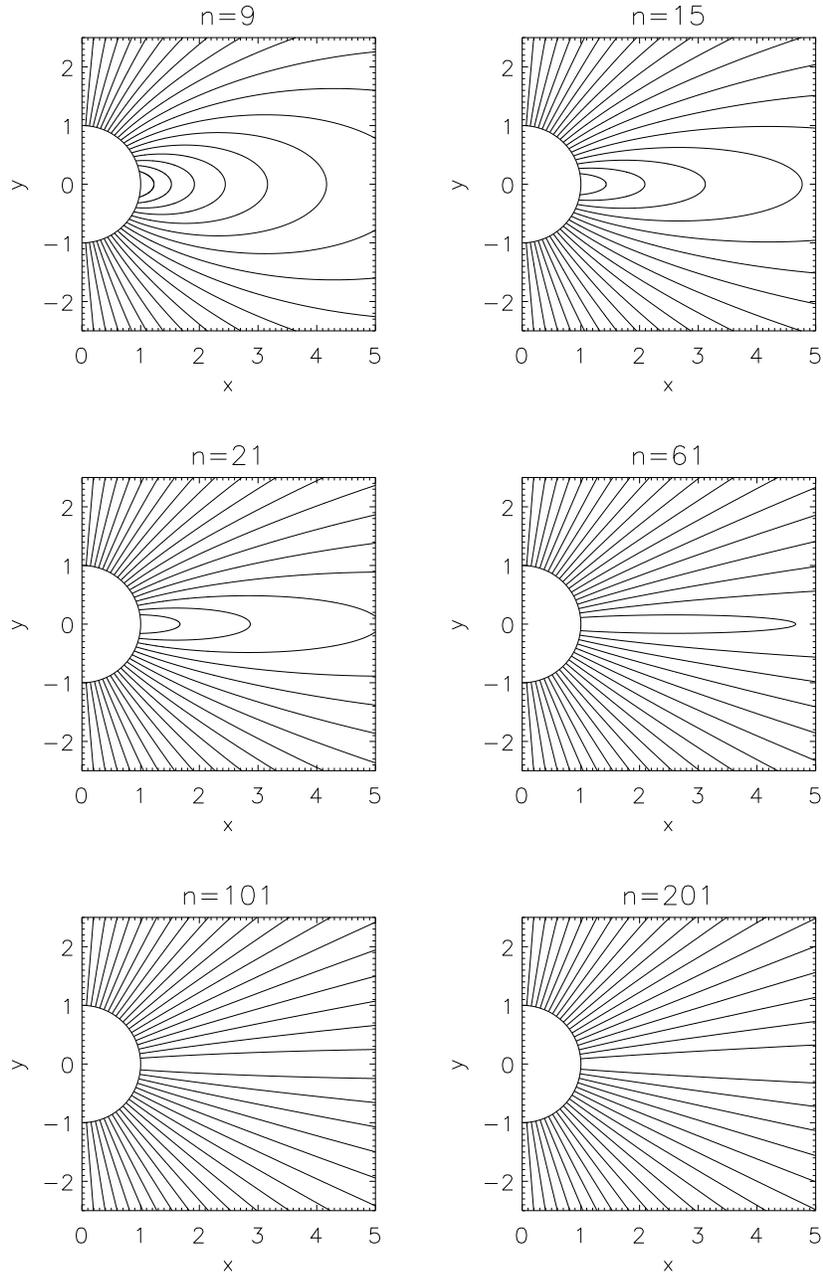}}
\caption{\small{Evolution of the field configuration of self-similar force-free fields. We see here that the field lines become more and more open as $n$ increases.}}
\end{figure}

\begin{figure}
\centerline{\includegraphics[width=130mm]{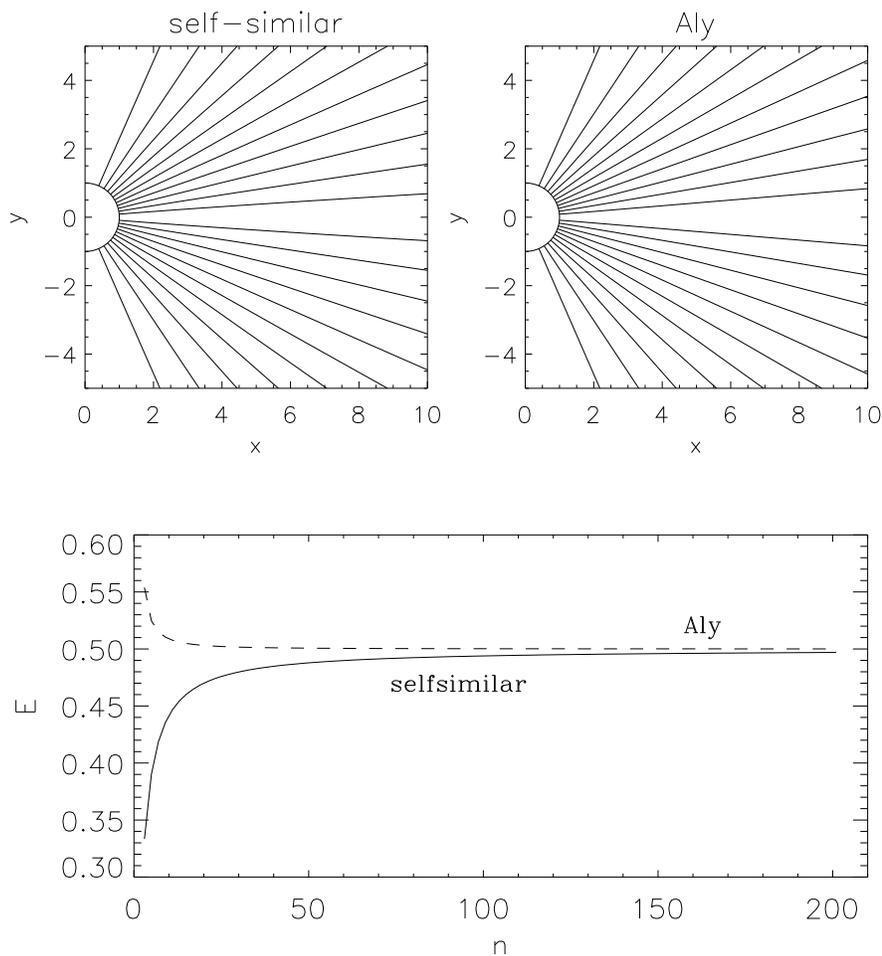}}
\caption{\small{Top panels: Field configurations of the $n=201$ self-similar force-free field (left) and the potential fully-open Aly field (right). We see here that the two fields show very similar configurations in the $r-\theta$ plane. Bottom: Variations of the magnetic energies of self-similar force-free fields (solid line) and the fully-open Aly fields (dashed line) with different $n$. We see here that as $n$ increases the magnetic energies of the self-similar force-free fields and the fully-open Aly fields approach one another.}}
\end{figure}

\begin{figure}
\centerline{\includegraphics[width=140mm]{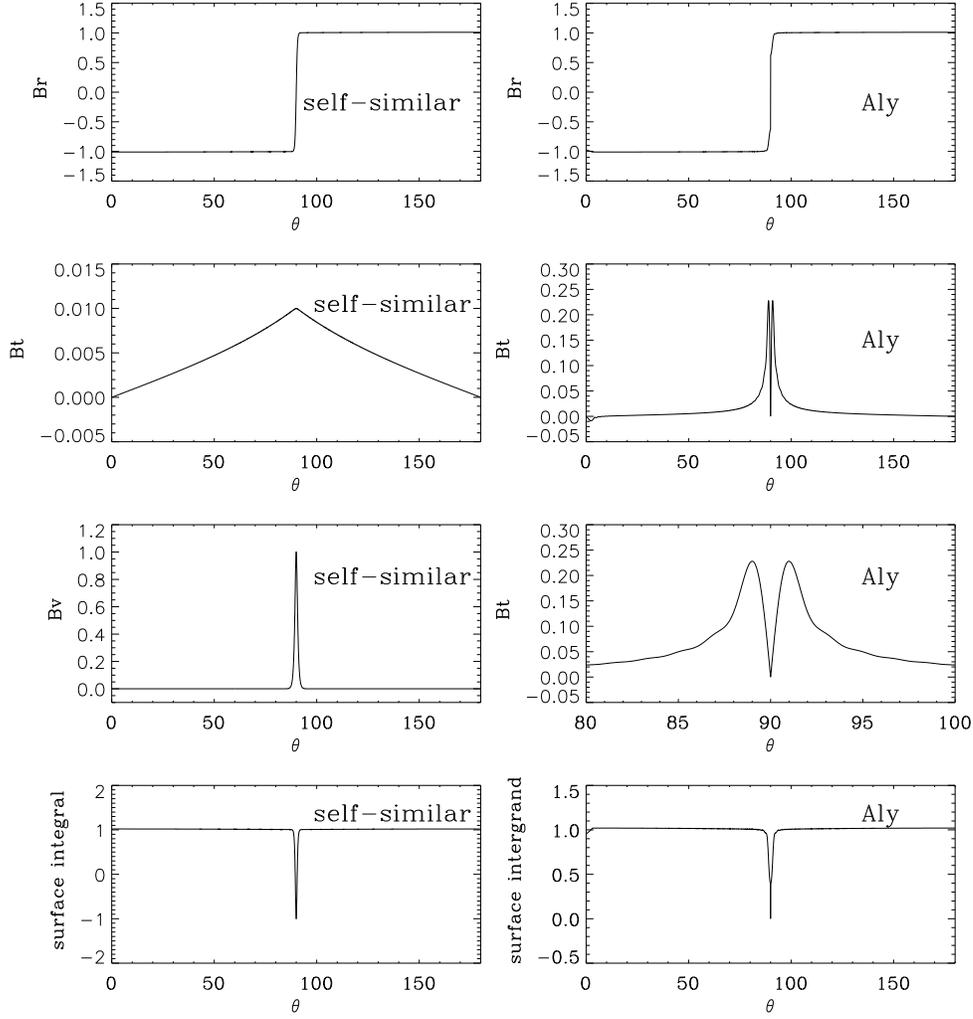}}
\caption{\small{A comparison of several physical quantities between the $n=201$ self-similar force-free field (left panels) and the potential fully-open Aly field (right panels). These physical quantities are, from top to bottom panels respectively, normal magnetic field ($B_r$), transverse magnetic field ($B_t$), azimuthal magnetic field ($B_v$) and surface integrand ($B_r^2-B_t^2-B_v^2$) at $r=1$ vs different latitudes. By definition, $B_v=0$ for Aly field. Shown in the third-right panel is the zoomed-in curve of $B_t$ in the second-right panel.}}
\end{figure}

\begin{figure}
\centerline{\includegraphics[width=140mm]{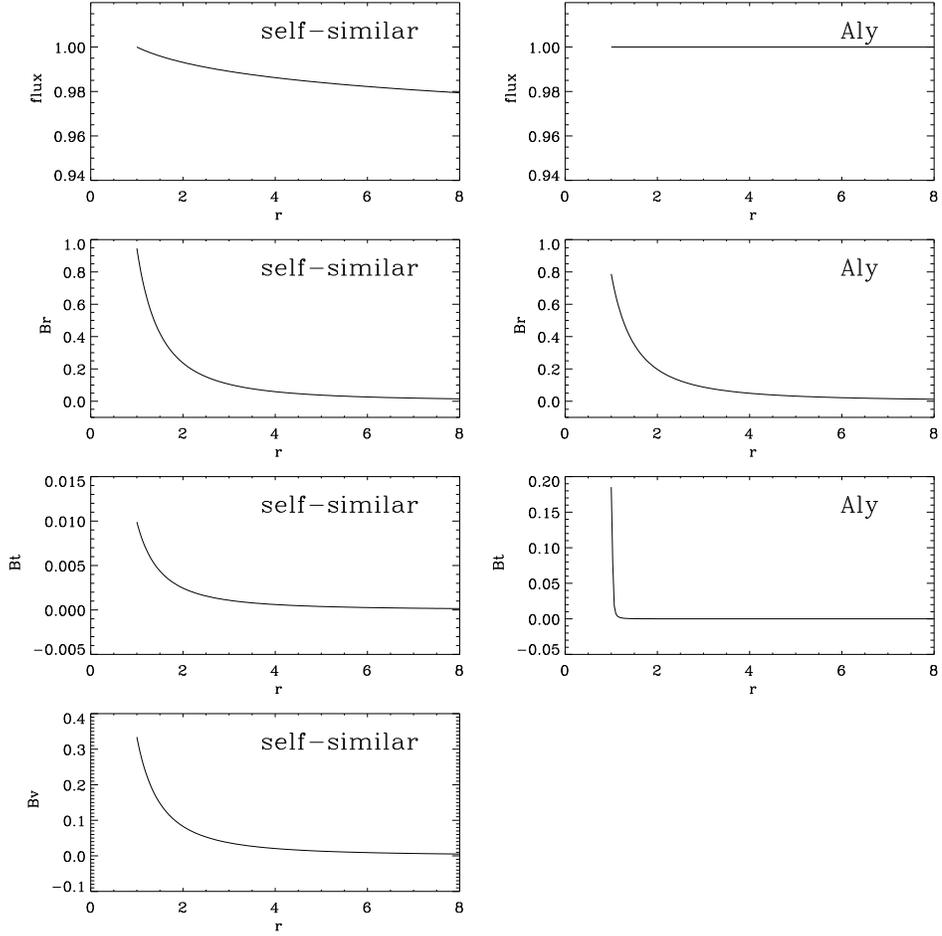}}
\caption{\small{Same as Figure 8, but now the comparison is made on the variations of flux function, normal magnetic field $B_r$, transverse magnetic field $B_t$ and azimuthal magnetic field $B_v$ at the equator, vs the radial distance.}}
\end{figure}

\begin{figure*}[tp]
\centering
\begin{minipage}[t]{0.4\textwidth}
{\includegraphics[width=0.75\textwidth]{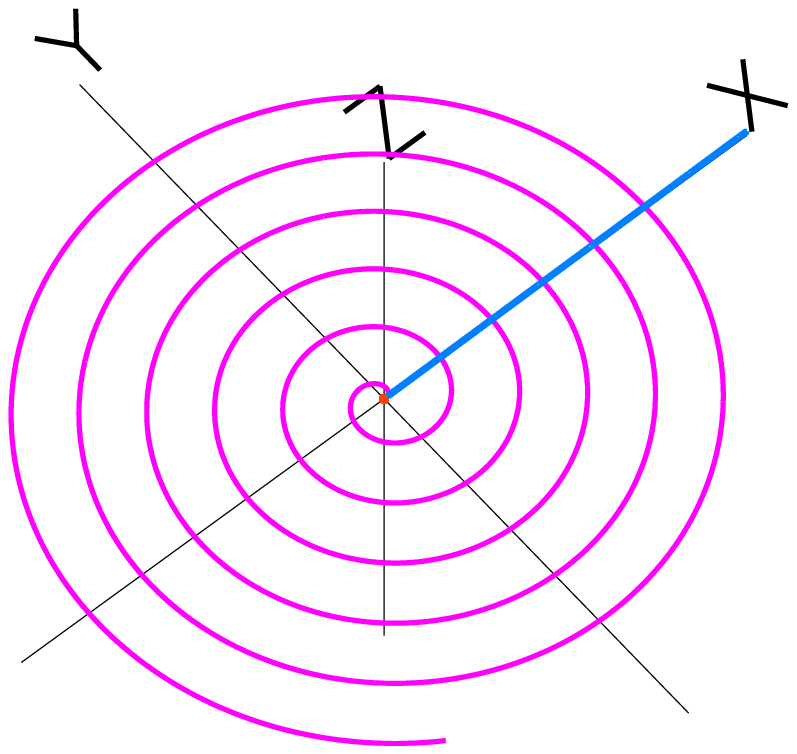}}
\end{minipage}
\begin{minipage}[t]{0.4\textwidth}
{\includegraphics[width=0.75\textwidth]{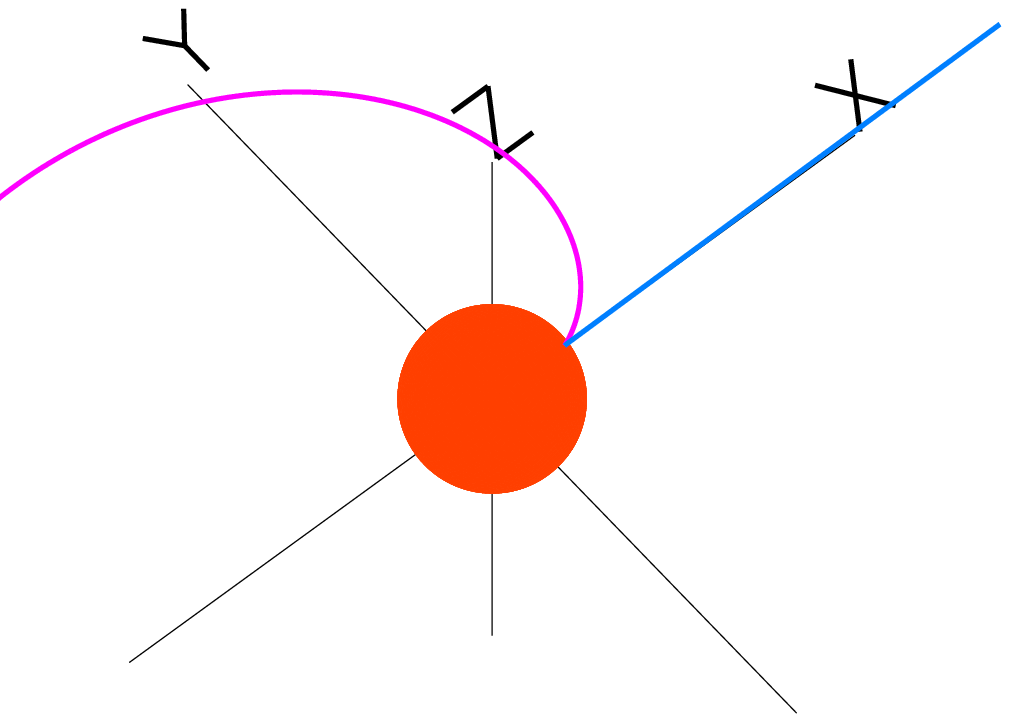}}
\end{minipage}
\begin{minipage}[t]{0.4\textwidth}
{\includegraphics[width=0.75\textwidth]{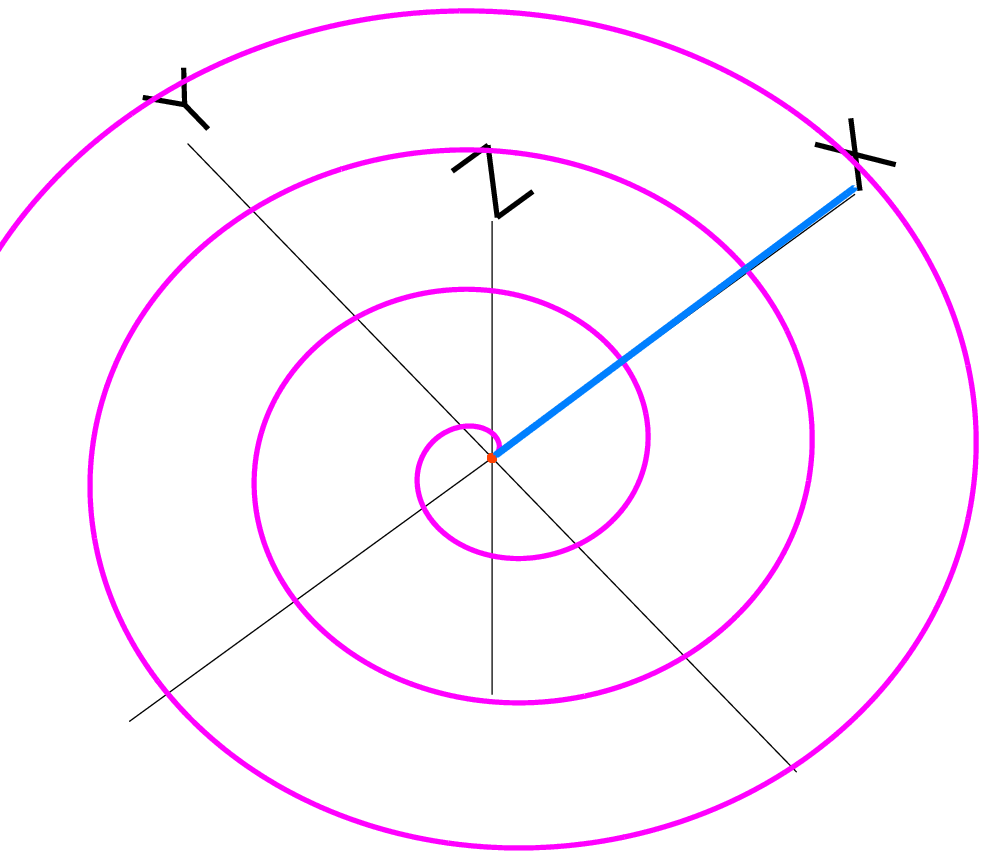}}
\end{minipage}
\begin{minipage}[t]{0.4\textwidth}
{\includegraphics[width=0.75\textwidth]{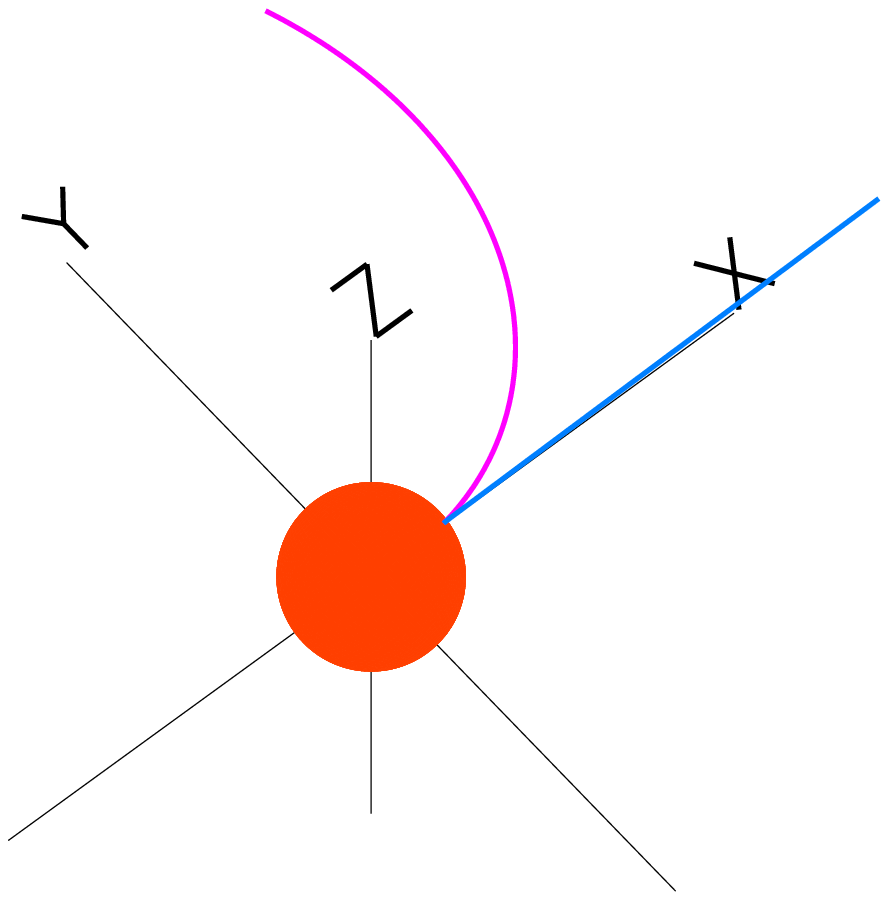}}
\end{minipage}
\begin{minipage}[t]{0.4\textwidth}
{\includegraphics[width=0.75\textwidth]{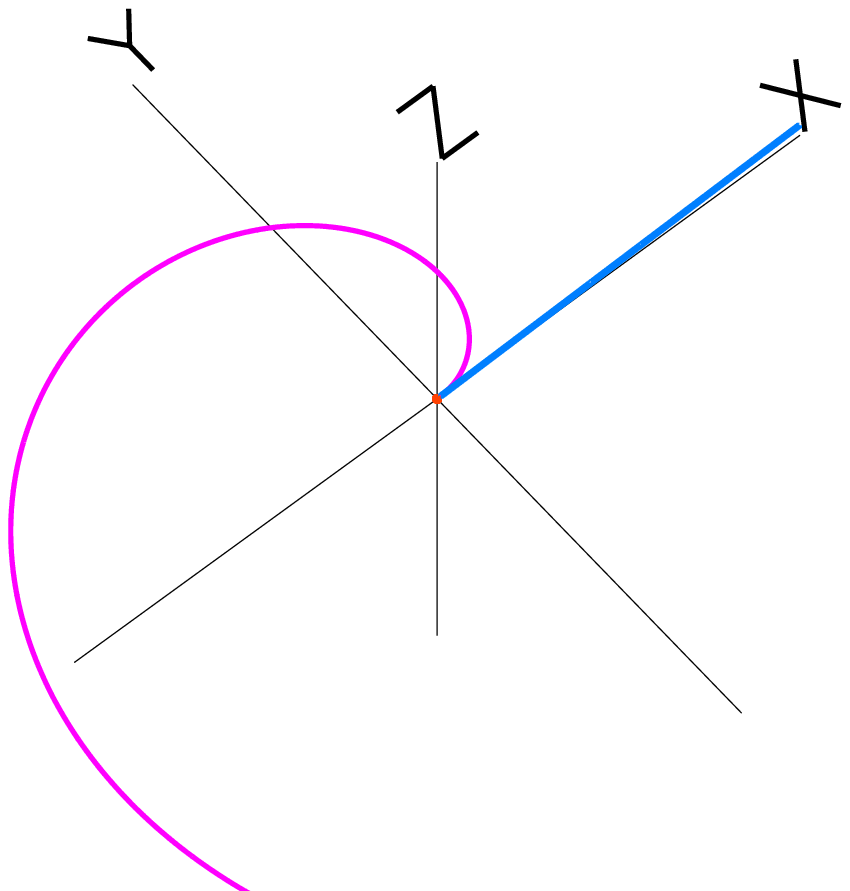}}
\end{minipage}
\begin{minipage}[t]{0.4\textwidth}
{\includegraphics[width=0.75\textwidth]{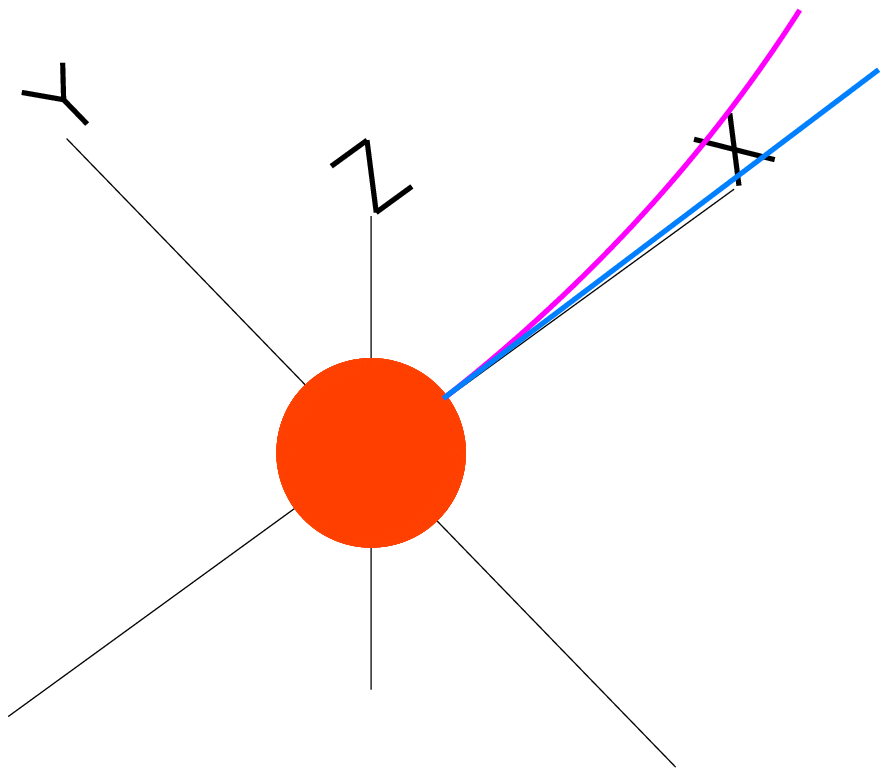}}
\end{minipage}
\begin{minipage}[t]{0.4\textwidth}
{\includegraphics[width=0.75\textwidth]{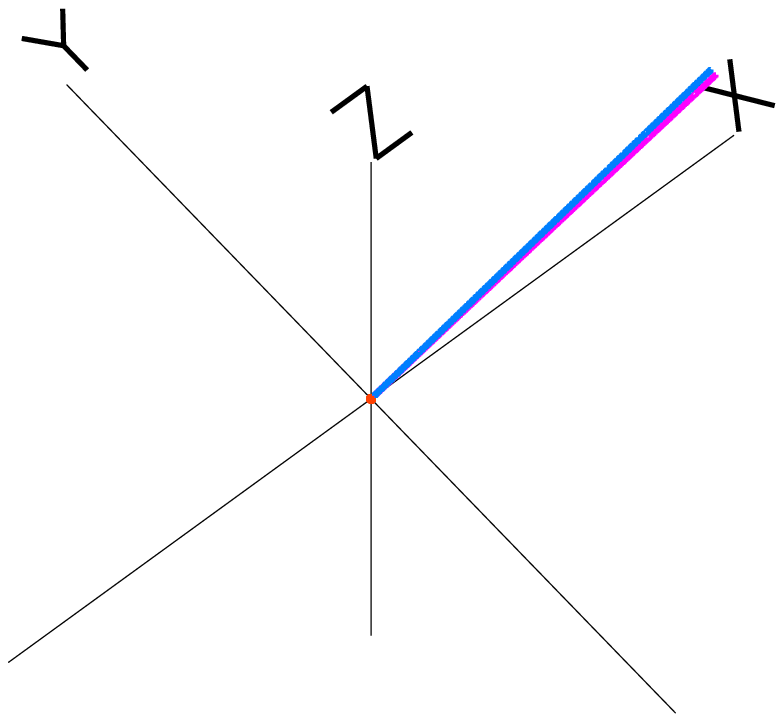}}
\end{minipage}
\begin{minipage}[t]{0.4\textwidth}
{\includegraphics[width=0.75\textwidth]{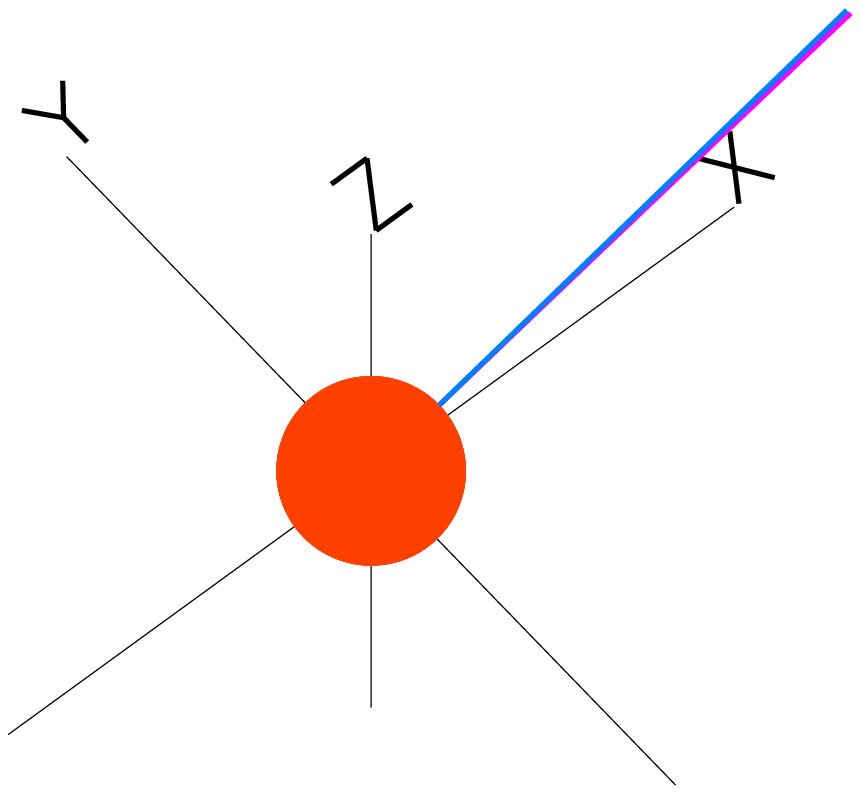}}
\end{minipage}
\begin{minipage}[b]{\textwidth}
\caption{\footnotesize{A comparison of the 3D structure of field lines between the $n=201$ self-similar force-free field (purple lines) and the fully-open Aly field (blue lines). From top to bottom panels, plotted respectively are the field lines $0.5^\circ$, $1^\circ$, $2^\circ$ and $20^\circ$ away from the equator. The equator is located in the X-Y plane and the poles in the Z direction. The length of each axis in the left panels is 100 solar radius and the length of each axis in the right panels is 5 solar radius. The central red sphere in each panel shows the size of the Sun.}}
\end{minipage}
\end{figure*}


\end{document}